\documentclass[sn-mathphys-ay]{sn-jnl}% Math and Physical Sciences Author Year Reference Style
\usepackage{graphicx}%
\usepackage{multirow}%
\usepackage{amsmath,amssymb,amsfonts}%
\usepackage{amsthm}%
\usepackage{mathrsfs}%
\usepackage[title]{appendix}%
\usepackage{xcolor}%
\usepackage{textcomp}%
\usepackage{manyfoot}%
\usepackage{booktabs}%
\usepackage{algorithm}%
\usepackage{algorithmicx}%
\usepackage{algpseudocode}%
\usepackage{listings}%
\usepackage{setspace}

\usepackage{subcaption}
\usepackage{epstopdf}
\begin{document}

\title[Dynamics of contaminant flow through porous media containing random adsorbers]{Dynamics of contaminant flow through porous media containing random adsorbers}


\author*[1]{\fnm{Kaj} \sur{Pettersson}}\email{kajp@chalmers.se}

\author[2]{\fnm{Albin} \sur{Nordlander}}

\author[2]{\fnm{Angela} \sur{Sasic Kalagasidis}}\email{Angela.Sasic@chalmers.se}

\author[2]{\fnm{Oskar} \sur{Modin}}\email{oskar.modin@chalmers.se}

\author[1]{\fnm{Dario} \sur{Maggiolo}}\email{maggiolo@chalmers.se}

\affil*[1]{\orgdiv{Department of Mechanics \& Maritime Sciences}, \orgname{Chalmers University of Technology}, \orgaddress{\street{Chalmersplatsen 4}, \city{Göteborg}, \postcode{412 96}, \country{Sweden}}}

\affil[2]{\orgdiv{Department of Architecture \& Civil Engineering}, \orgname{Chalmers University of Technology}, \orgaddress{\street{Chalmersplatsen 4}, \city{Göteborg}, \postcode{412 96}, \country{Sweden}}}

\abstract{Many porous media are mixtures of inert and reactive materials, manifesting spatio-chemical heterogeneity. We study the evolution of scalar transport in a chemically heterogeneous material that mimics a green roof soil substrate, fractionally composed of inert and reactive adsorbing particles. These adsorbing particles are equivalent to biochar within a real soil substrate. The scalar transport evolution is determined using experiments and simulations calibrated from experimental data. Experiment 1 is used to determine the equilibrium capacity and adsorption rate of two biochar types when immersed in a methylene blue solution. Breakthrough curves of a packed bed of glass beads with randomly interspersed biochar are determined in experiment 2. Simulations are then run to investigate the solute transport and adsorption dynamics at the pore-scale. An analytical model is proposed to capture the behavior of the biochar adsorption capacity and the simulation results are compared with experiment 2. A pore-scale analysis showed that uniformly sized beds are superior in contaminant breakthrough reduction, which is related to the adsorptive surface area and the rate at which adsorption capacity is reached. Cases using the adsorption capacity model display a tight distribution of particle surface concentration at later simulation times, indicating maximum possible adsorption. The beds with dissimilar particle sizes create more channeling effects which reduce adsorptive particle efficiency and consequently higher breakthrough concentration profiles. Comparison between experiments and simulations show good agreement. Improved biochar performance can be achieved by maintaining particle size uniformity alongside high adsorption capacity and adsorption rates appropriate to the rainfall intensity.}

\keywords{biochar, lattice Boltzmann, experiment, adsorption, methylene blue}

\maketitle

\section*{Article Highlights}
\begin{itemize}
    \item Experiments on biochar adsorption of methylene blue compared to simulations, with good agreement.
    \item Simple analytical model proposed to capture biochar adsorption capacity, results compared to experiments.
    \item Performance inferior in polydisperse beds due to underutilized particles and inhomogeneous concentration front profile.
\end{itemize}

\section{Introduction}\label{introduction}
Porous media containing a proportion of inert and chemically reactive elements can found in biology in the form of biofilters, bioreactors, and organic tissue. They are also found in geological elements such as rocks or soils, which are also highly chemically heterogeneous, containing a wide variety of chemically dissimilar minerals and/or organic elements. Research into this type of porous media has formerly been primarily driven by the petrochemical industry, however this has shifted in recent years to a more environmental focus. Green roofs are one such example, consisting of living greenery ranging from trees to grasses growing in a soil substrate whose composition can vary widely, from crushed brick to manure to peat moss. The benefits of green roofs in urban environments are well documented, such as their contributions to the reduction in urban noise, air, and water pollution; their effects on the urban heat island as well as building envelopes themselves; and urban rainwater runoff management \cite{AguilarFajardo20221937}. 

Green roofs require regular maintenance, including the administration of fertilizer upon installation and potentially thereafter as the need arises. The fertilizer consists of a fast-acting agent which gives the plants a boost to survive the initial period of installation and establish themselves; and longer-acting nutrients which aim to keep the additional required maintenance to a minimum. Excess fertilizer which cannot be stored by the soil or used directly by the vegetation can be carried by rainwater from the soil through the drainage system to locations where it acts as a contaminant \cite{Wang201765}. The addition of biochar to the soil to adsorb excess fertilizer (contaminant or solute) is currently being put forth as viable solution to this problem.

\subsection{Biochar in green roofs}
Biochar is essentially any organic material that has been carbonized under high temperature through a process known as pyrolysis and can vary in material properties and morphology significantly. A thorough overview of the types of biochar in use commercially can be found in \cite{Novotný2023}.

Biochar is incorporated into green roof soil via three strategies; random application to the top of the bedding material, thorough mixing into the substrate composition prior to vegetation, and set as a layer at the bottom of the substrate. Thorough mixing and layering at the soil base are considered the most effective, with the latter shown to be effective for reducing the leaching of total nitrogen and total phosphorous \cite{Kuoppamäki2016}. The proportion of applied biochar varies, from 5\% by weight/volume up to about 40\%, however excessive proportions of biochar will have adverse effects on plant growth and contribute to increased contamination \cite{XIANG2021}. 

There are few large-scale projects incorporating biochar use on green roofs though a few exist, such as the NWE CASCADE project, in which France intends to implement the use of biochar to enhance stormwater management within the Brittany region. Other similar solutions will likely be implemented within a larger initiative led by Bloomberg Philanthropies as touched upon in \cite{Senadheera2024}. 

The industrial-scale implementation is dependent upon several factors, such as the ecological impact of large-scale biochar use and the release of additional contaminants present within or on the surface of biochar as a result of its preparation and function. For example, \cite{Premarathna2023} discusses the almost complete removal of ammonia from the soil when excessive biochar is applied, which can be problematic for soil health. Another key factor is availability of biochar itself, given the different feedstock and treatment methods for its creation. A final major consideration is the lifetime of the biochar as an adsorbent and its end-of-life handling. Little work has been done on the effects of aging on biochar performance due to limited long-term projects but what is known is that degrading performance is dependent upon the contaminant and environmental conditions. For example, if the primary goal of the biochar is to trap micro and nano-plastics (MPs and NPs) then both the aging of the biochar \textit{and} the plastics themselves play an important role in the removal efficiency \cite{Ji2024}. Physical degradation of the biochar also occurs during the aging process and leads to biochar dust, which no longer serves its purpose of entrapping pollutants but can act to spread captured plastics or heavy metals due to its increased mobility. It has been shown the granulated biochar in particular is more resistant to this degradation process, thus making it more suitable for applications such as green roofs where the environmental exposure is high \cite{Lee2024}.

Related to the issue of biochar disposal is the issue of regeneration, which heavily depends upon the captured contaminants. It is possible to regenerate biochar that has captured MPs/NPs using organic solvents such as acetone, however this is not feasible in a green roof environment due to environmental concerns. Water rinsing performs relatively poorly for removing carbon-based adsorbents but can be used \cite{Ji2024}. Adsorbed copper can be almost completely recovered \cite{Bashir2023577}, as can cadmium \cite{Cui2022}. Biochar regeneration over five adsorption-desorption cycles was assessed with regard to volatile organic compounds (VOCs) with reported 88\% to 96\% regeneration \cite{Rajabi2021}. While these reports are promising, the scalability for some of the processes used for regeneration is in doubt as they are either inappropriate for green roofs due to their open nature, or less efficient if alternatives are used \cite{Gupta2020}.

Biochar may raise the production costs of green roofs, resulting in higher final prices for these products \cite{KHAN2021}. To evaluate its profitability for both producers and end users, it is essential to compare its effects against using fertilizers alone or in combination with fertilizers \cite{Ye2020}. Additionally, given its environmental benefits—such as minimizing nutrient leakage and boosting carbon sequestration—efforts are underway to improve cost calculation models to include both private and societal costs and benefits \cite{CAMPION2023}. However, these calculations are highly case-specific, influenced by factors like location, feedstock, scale, pyrolysis conditions, biochar pricing, and crop type. To promote greater adoption of biochar, \cite{CAMPION2023} suggest developing standardized calculation models and emphasizing additional societal benefits of biochar, such as improved water retention. While the proposed model is comprehensive, it remains unclear how potential hazards associated with biochar \cite{XIANG2021} will be addressed with it. Thus, the technoeconomic benefits of biochar will be a focus of future research.

\subsection{Biochar adsorption capabilities}
Many variants of biochar have been employed for the purpose of removing contaminants in a soil or packed bed. 
\cite{Afroze20162343} determined the adsorption of methylene blue (MB) by eucalyptus bark biomass in a packed bed and found that it performs well in the removal of dye-containing effluents. \cite{Dawood2018} analyzed the adsorption of MB in pine cone biochar in a packed bed column of Kaolin clay and utilized several analytical models to determine breakthrough curves. These curves were compared under varying experimental conditions and the biochar/clay packed bed was found to perform best under low flow rates, high MB concentration, and larger bed depth. \cite{ZANINLIMA2023118515} analyzed competitive sorption and desorption of zinc, cadmium, and lead between compost, biochar, and peat. They found that biochar had the highest adsorption capacity and lowest desorption rate, with lead removal being the most effective. \cite{Beesley20102282} examined the efficacy of biochar and greenwaste compost on the reduction of zinc, cadmium, and copper within soils and found them to be beneficial in the control of pollutants. \cite{PITA2024100596} provided a review of the work associated with the use of biomass on the removal of pharmaceutical compounds and reported that biochar made from a variety of plants; such as rice husks, corn, sawdust, and sugar cane, was the most effective in many industrial applications, partly due to their high adsorption capacity. While the aforementioned studies are valuable in understanding biochar as an adsorbent for a variety of contaminants, we are unable to identify the mechanisms by which it performs best, such as available adsorbing surface area, effect on the flow field, or their pore storage capacity. For this level of detail we must utilize numerical models which can describe such systems and provide the level of detail we require. 

The physical process of contaminant adsorption in a flow field is described mathematically by the advection-diffusion of species, combined with adsorption/desorption; a model which is present across many fields of application. This general model is used to describe processes such as post-combustion CO$_2$ capture \cite{Proll2016166}, determining optimal geometry of granular heat exchangers \cite{Mitra2018764}, analyzing methane adsorption in subsurface shale deposits \cite{Li2016675}, and describing the behavior of photocatalytic textiles \cite{Robin2016269} to name a few. Investigations into the dominant factors affecting adsorption rates and capacity have shown that particle and pore morphology play an important role \cite{Liapis199913,JARETEG2022682}. Adsorption rates themselves also strongly determine the process evolution \cite{Marin20142356}, with lower rates determining adsorption quantity by reaction duration, and higher rates controlling quantity through reaction rate \cite{Zakirov2023}. Non-isothermal systems can result in high heterogeneity of concentration and temperature, with performance suffering from regions of stagnant flow and increased near-wall breakthrough \cite{Verma20091003}.

When one considers a transported reactive species adsorbed locally by reactive elements within a porous matrix; understanding the underlying interactions which determine the local concentration distribution is necessary. The mixing itself within porous media has been shown to be complex at high Reynolds numbers (Re). As such, the flow is characterized by an initial advection-dominated regime as the flow penetrates the pore network and is followed by a diffusion-dominated regime wherein the molecular diffusion redistributes the scalar more effectively than the advection. The advective regime consists of stretching of the scalar front in longer finger-like structures, particularly where the flow is strongest, and can be defined by by the plume’s concentration mean, variance, and probability distribution \cite{Bonazzi2023369}. A thorough overview of current state of research and methods used to solve species transport in porous media can be found in \cite{Dentz20231}.

At low Re, over a short time these structures regress to a more uniform distribution through molecular diffusion. However, even at low Re, the process of homogenization by diffusion may be retarded by the presence of local chemical heterogeneities in the form of solid adsorbing particles which affect the concentration field. The effect of varying the quantity of uniformly distributed adsorbers within a porous medium was undertaken previously, with two regimes in the evolution of the concentration field identified and the rate at which one regime reaches the other is determined by the number of adsorbing elements present \cite{Maggiolo2023}. 

The previous investigations cover primarily changes in adsorptivity, usually a static value, with the values not necessarily reflecting realistic conditions. The geometries generally tend to be regular, not reflecting in any way a realistic soil sample. We will extend this characterization by introducing the presence of polydisperse particles within the packed bed as well as an adsorption rate dependent upon a proposed adsorption capacity model for the biochar. In addition, we make use of experimental data to correctly tune the simulations to realistic adsorption rates and determine how these factors impact the previously identified regimes as well as report additional influences on the biochar performance. The results are directly compared to experimental data to assess the predictive accuracy of our model, quantified by breakthrough curves.

\section{Experiments}
The experiments outlined in this section have been undertaken for two purposes. Experiment 1 was used calculate the adsorption capacity of the biochar which in turn is an input in the numerical simulations used to compare against experiment 2. The results for experiment 2 were directly compared to the numerical results to determine the accuracy of the model used in this work. All experimental information is given in this section, with the common materials introduced first, followed by each experiment and its respective results.

\begin{figure}[!ht]
   \centering
    \includegraphics[width=0.6\textwidth]{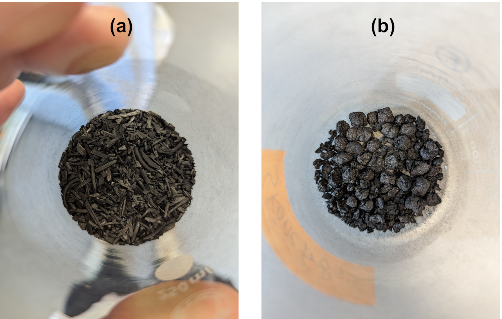}  
    \caption{Types of biochar (a) WCB (Hjelmsäters Fastigheter AB) and (b) GB (Skånefrö AB).}\label{fig:fig1}
\end{figure}

The adsorption capabilities of two types of biochar were experimentally determined: wood-chip biochar (WCB) produced mainly from European spruce and granulated biochar (GB) produced mainly from agricultural seed waste. The wood chip biochar was produced by Hjelmsäters Fastigheter AB while the granulated biochar was produced by Skånefrö AB; both companies being based in Sweden. Figure \ref{fig:fig1} displays the types of biochar used in the experiments. Alongside the types of biochar, acid-washed glass beads were also used and acted both as an inert, transparent packing material for the biochar and as a control case with minimal to no adsorption.

MB, chemical formula C$_{16}$H$_{18}$CIN$_3$S, was used as the adsorbate; diluted to 6 mg/L in a phosphate buffer solution containing 380 mg/L KH$_2$PO$_4$ and 300 mg/L K$_2$HPO$_4$. It is worth noting that due to its properties, methylene blue can be seen as an analog for some organic pollutants however it is by no means representative of all possibilities. Other compounds or the direct pollutant itself may be required for more accuracy and the numerical model must be appropriately modified as well. 

\begin{figure}[!ht]
   \centering
    \includegraphics[width=0.8\textwidth]{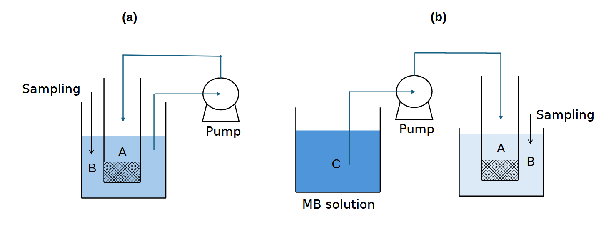}      
    \caption{Experimental setup for (a) experiment 1 and (b) experiment 2. A - Inner annular column containing porous medium, B - Outer annular beaker where MB sampling occurs, C - MB reservoir with original concentration $c_0$ used in experiment 2}
    \label{fig:fig2}
\end{figure}

The MB concentration was measured using spectrophotometry with absorbance measured at 680nm. Figure \ref{fig:fig2} shows the experimental setup that was used for experiments 1 and 2.

\subsection{Experiment 1}
\begin{figure}[!ht]
\centering
    \includegraphics[width=0.6\textwidth]{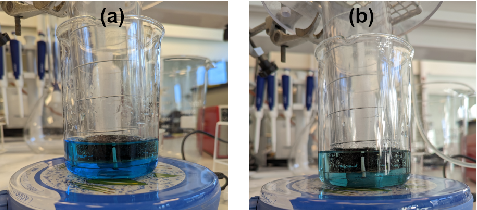}        
\caption{Experiment 1 at (a) $t=0$ min and (b) $t=135$ min}
\label{fig:fig3}
\end{figure}

In experiment 1, a known mass of biochar or glass beads (1mm diameter) was placed in a 5cm diameter Plexiglas column, as shown in Figure \ref{fig:fig3}. The quantities of biochar were chosen such that differences in behavior would be evident while allowing for a steady-state to be reached within a reasonable time frame.

The bottom of the column was covered with a steel mesh which kept the biochar/glass beads within the column. The column was submerged in a glass beaker containing 100mL of the MB solution, which was circulated through the biochar-packed column using a peristaltic pump operating at a flow rate of 4mL/min. The column is fully immersed in the MB solution to allow for the adsorption to occur at the maximum possible rate, which is a property of the biochar itself rather than a consequence of the local availability of adsorbent. Liquid samples from the solution being circulated through the Plexiglas column were collected regularly from the glass beaker. The quantity of biochar used in this experiment is determined such that the measured difference in adsorbance is clear as the quantity of biochar is increased. 

\begin{figure}[!ht]
\centering
\includegraphics[width=\textwidth]{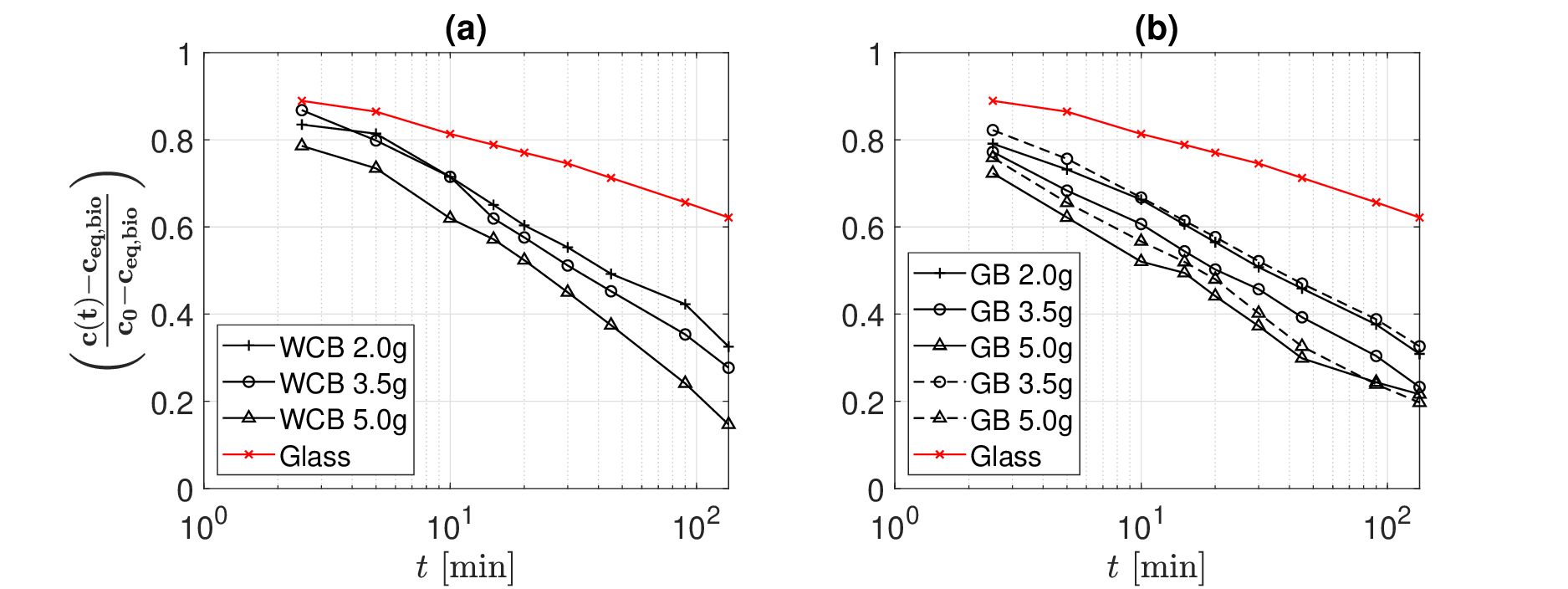}        
\caption{Normalized outflow concentration for (a) Wood chip biochar (WCB), (b) Granulated biochar (GB)}
\label{fig:fig4}
\end{figure}

Figure \ref{fig:fig4} shows the evolution of the normalized outflow concentration over time in experiment 1. Absorbance (A) is converted to concentration and normalized against the initial concentration $c_0$. We define absorbance as $A = \text{log}_{10}(I_0/I)$ where $I_0$ is the intensity of incident light at 680nm wavelength and $I$ is the transmitted intensity.
There is no large difference in the performance for the cases using WCB, even when the quantity of biochar is increased to 5 grams; with the results for the GB being similar. To calculate the adsorption capacity of the biochar we employ the conservation equation 
\begin{equation}\label{eq:eq1}
    V_{bio}c_{eq,bio}  = V_{liq}(c_0 - c(t_{end})) - V_{glass}c_{eq,glass}, 
\end{equation}
where $V_{bio},V_{glass}$ and $V_{liq}$ are the biochar, glass and liquid volumes, $c_0$ is the initial concentration, $c$ is the measured concentration, and $c_{eq,bio},c_{eq,glass}$ are the measured adsorption capacities (equilibrium concentrations) within the biochar and glass, which is the quantity at which they cease to adsorb addition solute, at $t_{end}$. Table \ref{tab:tab1} gives the substrate composition for experiment 1 and the resultant calculated adsorption capacity (equilibrium concentration).
\begin{table}[ht!]
	\centering
		\caption{Experiment 1, Q = 4mL/min MB solution. WCB - wood-chip biochar, GB - granulated biochar. Calculated adsorption rate $k=3.3e-4$. [s$^{-1}$]}\label{tab:tab1}
\begin{tabular}{ |c|c|ccc|ccccc| }
  \hline
  Material & Glass & & WCB & &  &  & GB &  &  \\ 
  \hline
  Weight [g] & 20.0 & 2.0 & 3.5 & 5.0 & 2.0 & 3.5 & 3.5 & 5.0 & 5.0 \\
  \hline
$c_{eq}$ & 0.24 & 0.42 & 0.45 & 0.52 & 0.43 & 0.47 & 0.42 & 0.48 & 0.49 \\
  \hline
\end{tabular}
\end{table}

Equation (\ref{eq:eq1}) is ideal for calculating the maximum adsorption capacity and biochar equilibrium concentration given that experiment 1 runs until a steady-state is reached, however this equation is unsuitable for calculating the evolution of the concentration at the adsorbing surface over time. To this end we employ the formulation for the evolution of the concentration at the adsorber surface
\begin{eqnarray}
    -V_p \frac{dc}{dt} &=& S_pk(c(t) - c_{eq,bio}),\label{eq:eq2}\\
    c(0) &=& c_0, \nonumber \\
    c(t\rightarrow \infty) &=& c_{eq,bio}, \nonumber
\end{eqnarray}
where the solution takes the form
\begin{equation}\label{eq:eq3}
    c(t) = (c_0 - c_{eq,bio})e^{-S_pkt/V_p} + c_{eq,bio},
\end{equation}
where $V_p,S_p$ are the reactive particle volume and surface area and the adsorption rate is calculated by solving (\ref{eq:eq3}) for $k$, as all other quantities are known. This formulation is based upon the assumption that the system is reaction-limited, which is valid since the biochar is initially fully immersed in the solution. Fully immersed biochar allows for the entire surface area of a reactive particle to contribute in the removal of the MB with maximum efficiency, hence reaction-limited. The calculated adsorption rate is used in the calibration of the simulations, as explained in section \ref{sec:num1}. It is important to mention that the above formulations cannot be used for the analysis of experiment 2 as it is mass-transport limited and thus we cannot make any assumption on the evolution of adsorption in time. Thus we are forced to model the rate at which the adsorption capacity of the biochar is reached, and it's subsequent effect on the adsorption rate itself. 

If one wishes to apply a simple capacity model to the system, the resulting boundary condition at the reactive particle surface will be as in equation (\ref{eq:eq4}). Note that this condition bears striking similarity to (\ref{eq:eq2}) but is distinct, supporting our claim equation (\ref{eq:eq3}) is not suitable for solving the mass-transport limited problem. In these equations $D_m$ is the molecular diffusivity, Da the Damköhler number, and $f(c)$ the capacity model. This condition will be formalized in section \ref{sec:num1} and is only presented here to demonstrate the differences in boundary condition between the experiments. 
\begin{eqnarray}\label{eq:eq4}
 -V_p \frac{dc}{dt} &=& S_pD_m \frac{\partial c}{\partial n}, \\
 \frac{S_pD_m}{Q} \frac{\partial c}{\partial n} &=& \frac{k S_\zeta}{Q} f(c) c(t),\nonumber \\
 &=& \text{Da } f(c)c^*|_{S_\zeta}.
\end{eqnarray}
\subsection{Experiment 2}
\begin{figure}[!ht]
    \centering
    \includegraphics[width=0.5\textwidth]{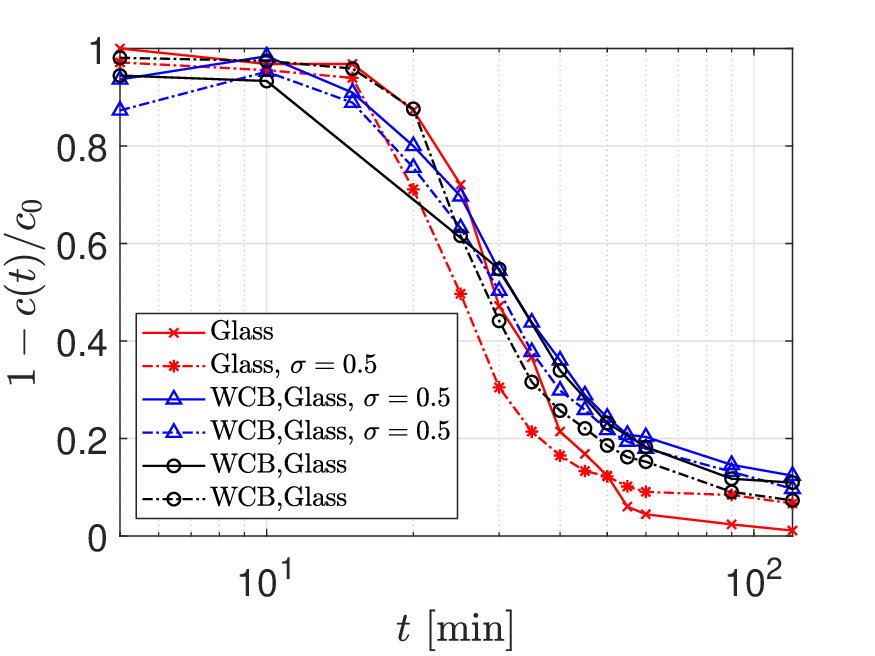}
    \caption{Evolution of concentration breakthrough $1-c(t)/c_0$}
    \label{fig:fig5}
\end{figure}

In experiment 2, the Plexiglass column was filled with glass beads mixed with WCB to a height of 32mm. The biochar made up 15\% of the column by volume and was either evenly mixed with the glass beads, placed as a layer at the bottom of the column, or placed as a layer at the top of the glass beads. The chosen fraction of 15\% is motivated by the standard composition used in green roofs in industrial production, which varies but provides diminishing benefits beyond this fraction. The column was submerged in a beaker filled with approximately 100mL phosphate solution without MB. The MB solution was then fed to the top of the column at a flow rate of 8mL/min and the MB concentration in the solution surrounding the column was measured in regular time intervals. Table \ref{tab:tab2} outlines the cases shown in Figure \ref{fig:fig5}, including variable particle distributions and the equivalent weight of the included biochar.

\begin{table}[ht!]
	\centering
		\caption{Experiment 2, Q = 8mL/min with MB solution. WCB - wood-chip biochar.}\label{tab:tab2}
\begin{tabular}{ |c|c|c|c| }
  \hline
  Material Type & \%Biochar & Biochar [g] & Particle dist. [mm] \\
  \hline
  Glass Only & 0.0 & 0.0 & $\mu = 1.0$ \\
  Glass Only & 0.0 & 0.0 & $\mu = 1.0, \sigma = 0.5$ \\
  \hline
  Glass + WCB & 15.0 & 0.93 & $\mu = 1.0, \sigma = 0.5$ \\
  Glass + WCB & 15.0 & 0.93 & $\mu = 1.0, \sigma = 0.5$ \\
  Glass + WCB & 15.0 & 0.93 & $\mu = 1.0$ \\
  Glass + WCB & 15.0 & 0.93 & $\mu = 1.0$ \\ 
  \hline
\end{tabular}
\end{table}

The results of experiment 2 are shown in Figure \ref{fig:fig5} in the form of normalized concentration outflow, also called breakthrough. Once again the data is normalized against an initial concentration.
We see an initial period where little of the MB is adsorbed, whereafter the reaction rapidly reaches saturation within the biochar and little additional MB solution is adsorbed, leading to a "steady state" in the breakthrough curve. These periods; the initial adsorption, the rapid saturation, and the ceasing of additional adsorption can be characterized by the time intervals of i) $0<t<20$, ii) $20<t<70$, and iii) $t>70$. 

We see that the cases using only glass beads perform the worst, as expected. Interestingly these cases also adsorb some measure of MB, which means their behavior should be taken into account when examining the results of the cases which include biochar. Indeed, the inclusion of the quantities of biochar used in these experiments show that they perform only marginally better than the glass beads themselves. Another important point to note from Figure \ref{fig:fig5} is that there is little discernible difference between cases in which a variety of particle sizes are present and cases where only a single particle size are used. This will be touched upon later in the numerical simulation results, particularly when the concept of adsorption capacity of reactive particles is present within the system.

\section{Numerical approach}
\subsection{Lattice Boltzmann method}
When one talks about solving the transport of a scalar, we formalize the concept using the advection-diffusion equation
\begin{equation}
    \frac{\partial c}{\partial t} = \nabla \cdot (D_m\nabla c) - \nabla \cdot (\mathbf{u}c) + R,
\end{equation}
where $c$ is a scalar (in our case concentration), $t$ is time, $D_m$ is the diffusion coefficient, $\mathbf{u}$ is the advecting fluid velocity, and $R$ is a sink/source term related to reactive processes. At the micro-scale this equation can be represented using the Boltzmann transport equation, which describes motion through particle streaming as well as collisions.

It is an ideal choice for solving flows in porous media due to the complex geometry involved and allows for detailed information of the flow dynamics to be extracted at the pore scale. The system is solved on a lattice structure wherein each lattice element consists of a centroid and nodes placed on a cubic convex hull. The fictive particles travel along the lattice nodes governed by probabilities appropriate for the chosen lattice geometry such that the macroscopic properties of the fluid are preserved \cite{Succi2001}. A 3D regular cubic lattice with 19 degrees of freedom for movement (D3Q19) is used and the solved equation is of the form
\begin{equation}
f_r (\mathbf{x} + \mathbf{c}_r \delta t, t+\delta t) - f_r(\mathbf{x},t) =  - \tau^{-1} ( f_r(\mathbf{x},t)-f^{eq}_r (\mathbf{x},t)  ) + F_r 
\end{equation}
where $f_r(\mathbf{x},t)$ is the distribution function at position $\mathbf{x}$ and time $t$ along the $r$-th direction; $\mathbf{c}_r$ is the so-called discrete velocity vector along the $r$-th direction over time interval $\delta t$; $f_r^{eq}$ is the equilibrium distribution function; and $\tau$ is the mean collision time and is related to kinematic viscosity by $\nu = c_s^2(\tau - 0.5 \delta t)$. The equilibrium distribution function $f_r^{eq}(\mathbf{x},t)$ takes the form
\begin{eqnarray}
f^{eq}_r &=& w_r \rho\ \Bigg ( 1 - \frac{\mathbf{u}\cdot\mathbf{u}}{2c_s^2} \Bigg ) \ , \ r=1 \\
f^{eq}_r &=& w_r \rho\ \Bigg (1+\frac{\mathbf{c}_r \cdot \mathbf{u}}{c_s^2} + \frac{(\mathbf{c}_r \cdot \mathbf{u})^2}{2c_s^4} - \frac{\mathbf{u}\cdot \mathbf{u}}{2c_s^2} \Bigg ) \ , \ r=2-19  
\end{eqnarray}
where $w_r$ is the appropriate weighting parameter for the D3Q19 lattice; $\rho$ is the density; $c_s$ is the speed of sound; and $\mathbf{u}$ is the velocity used for defining the equilibrium distribution functions, which can differ from the fluid hydrodynamic velocity, on the basis of the specific forcing scheme used.

The macroscopic flow quantities density and velocity, $(\rho,\mathbf{u})$ are thus related to the hydrodynamic moments as the following:
\begin{eqnarray}
\rho &=& \sum_r f_r \ , \\
\rho \mathbf{u} &=& \sum_r \mathbf{c}_r f_r + \frac{\Delta t}{2} \left( \frac{\Delta P}{L}\right),
\end{eqnarray}
where $\Delta t = 1$ in our case. The force $F_r$ as formulated by \cite{Guo2002} is applied to the fluid, which mimics the flow rate intensity during a rain event in our case as described later in more detail, and is given by
\begin{eqnarray}
F_r = \Bigg ( 1-\frac{1}{2\tau} \Bigg ) w_r \Bigg ( \frac{\mathbf{c}_r-\mathbf{u}}{c_s^2} +\frac{\mathbf{c}_r\cdot\mathbf{u}}{c_s^4}\mathbf{c}_r \Bigg )\left( \frac{\Delta P}{L}\right).
\end{eqnarray}

\subsection{Solute adsorption implementation}\label{sec:num1}
A solute adsorption rate is assigned to a percentage of chosen particles $\zeta$ such that their surfaces $S$ follow the first-order kinetics given by:
\begin{equation}
    - \frac{\partial c^*}{\partial \lambda_s^*}|_{S_\zeta} = \text{Da } c^*|_{S_\zeta} ,
\end{equation}
where Da is the Damköhler number, $c^* = c(\mathbf{x},t)/c_0$ is the dimensionless concentration at position $\mathbf{x}=(x,y,z)$, and $\lambda_S^* = \lambda_S/d$ is the dimensionless direction pointing inward to the particle surface. The Damköhler number, a ratio of the reaction rate and the advective mass flow rate, is given by
\begin{equation}
    \text{Da} = kS_{\zeta}/Q,
\end{equation} 
where $k$ is the adsorption rate on the particle surface, $S_{\zeta}$ is the total reactive surface area expressed as a fraction $\zeta$ of the total particle surface area, and the flow rate is given by $Q = l_0^2\varepsilon U$. Here $l_0$ is the porous domain width, $\varepsilon$ is the porosity, and $U$ is the mean streamwise velocity.

Numerically, the concentration field $c(\mathbf{x},t)$ is solved using a second population which is transported by the fluid velocity $\mathbf{u}$. The solute is then initialized at the inlet of the porous media as a volume with concentration $c_0$. The second lattice population, denoted $g_r$, gives the local concentration field by
\begin{equation}
    c(\mathbf{x},t) = \sum_r g_r(\mathbf{x},t).
\end{equation}

We impose a Neumann boundary condition at the adsorbing surfaces for the scalar lattice Boltzmann quantity $g_r$. The distribution function at a fluid node $\mathbf{x}$ in the proximity of an adsorbing surface placed at $(\mathbf{x} - \mathbf{c}_r)$ is corrected along the wall-normal direction r as
\begin{equation}
    g_r(\mathbf{x},t+1) = \frac{-A_1 + A_2}{A_1 + A_2}g_r(\mathbf{x},t) + \frac{2w_rA_3}{A_1 + A_2},
\end{equation}
where $A_1 = k$, $A_2 = D_m$, and $A_3 = 0$. This condition was developed by \cite{HUANG201570,HUANG201626}.

In order to limit the reactivity as the particle surface reaches high levels of concentration we apply a functional coefficient to the prescribed reaction rate. This limit is designed to act as an analog to the particles reaching a saturated state wherein they can no longer adsorb more of the solute, known as adsorption capacity. Several different functions were tested, including a triangular function as well as a sigmoid function however a simple linear function was chosen, of the form
\begin{equation}
    f(c) = 1 - c/c_{eq} \text{,    } \forall c \leq c_{eq}
\end{equation}
which allows for full reactivity at low values of concentration but reaches zero when the particle equilibrium concentration has been reached. This equilibrium concentration is set to $c_{eq} = 0.5$ and is based upon the results of experiment 1. This alteration is reflected in the adsorption boundary condition, becoming
\begin{equation}
    - \frac{\partial c^*}{\partial \lambda_s^*}|_{S_\zeta} = \text{Da } f(c) c^*|_{S_\zeta},
\end{equation}
which is the equivalent of equation (\ref{eq:eq4}) presented in the previous section.

\subsection{Domain generation and operating conditions}
The pore structure used in the simulations is a packed bed generated by the application of rigid body physics on falling spheres within a cylindrical container. Two such cases are generated; i) a monodisperse case with a fixed particle diameter, and ii) a polydisperse case with the mean diameter equal to that of the monodisperse case. The particles are centered around a diameter of $1.0$ with $2\sigma = \pm 0.82$ millimeters. A cubic subdomain is then selected from the cylindrical container and the volume is discretized into a binary matrix of $l_0^3 = 256^3$ computational nodes, giving the mean particle diameter $d=21.6$ nodes. The method by which the domain is generated and characterized can be found in \cite{Maggiolo2023} such that domain integrity is assured and the volumetric porosity is calculated to be $\varepsilon = 0.4, 0.36$ for the mono- and polydisperse cases, respectively. 

In order to simulate an intense rainfall event the Péclet number Pe $> 1$ and we assume little inertial influence, thus Re $< 1$. We set the Damköhler number, $\text{Da }\approx 1$, which represents equal magnitudes of molecular diffusion and reaction rates, as calculated in the experiment 2. By applying the conditions required above, the simulation has Re $=0.017$ and Pe $=6.7$, which is representative of a rainfall event with intensity of around 60mm/hr. 
The flow rate of experiment 2 is $Q = 8$mL/min, which converts to about 600mm/hr, about double the world record rainfall intensity \cite{WMO1947}, with Pe $= 250$ and Re $=0.17$. It is worthwhile to note that intensities higher than the record can be reached when runoff is channeled to a porous surface, which is not naturally occurring rainfall. While these values are higher than those chosen for the simulation, the governing physics are identically balanced, as the originally defined conditions are all satisfied, including Da $\approx 1$. With the above mentioned flow rates we have been able to measure variations in concentration quantities in a few hours, which we found as a good time lapse to minimize effects from changes in environmental conditions on the measurements.

These conditions are used to tune $\Delta P/L$, the molecular diffusivity and viscosity accordingly. It is worth mentioning that common rainfalls are between 3-8mm/hr, however less common extreme events can reach 15-60 mm/hr or more. We can extrapolate the simulation conditions to lower rainfall intensities by accounting for the time scale of the physical processes involved. The soil will not be as thin as that in the simulations and no runoff infiltration front will be pure contaminant. Additionally, one can alter the diffusivity to compensate such that the physically realistic process is retained.

The resultant mean flow velocity is averaged over the entire domain, given by $U=(\int_{l_0^3}u_{z}\text{d}l_0^3)/l_0^3\varepsilon$. The Kozeny-Carman relationship stipulates that if the porosity of a domain is altered the flow rate must be  changed accordingly. Since in our cases porosity is approximately unaltered, the resultant mean velocity for the polydisperse case is the same. The biochar percentage is set to 30\% in the simulations, double the industry maximum recommendation, to improve statistical analysis on single particle adsorption features without straying significantly from realistic conditions.   

The initial concentration $c_0 = 1.0$ is applied to an inlet zone directly adjacent to porous volume. The streamwise boundaries are assigned a periodic condition wherein buffer zones are applied to the inlet and outlet boundaries to prevent any effect on the flow within the porous zone. The lateral boundaries are assigned a free-slip condition. 

The simulations are run in two steps; first the single phase velocity profile is solved for each geometry and then the solute is added. In this way we solve only the steady-state flow in the first stage and only the concentration profile in time in the second stage. Six cases total are run on two different geometries, the details of which can be found in Table \ref{tab:tab3}.
\begin{table}[ht!]
	\centering
		\caption{Case numbers corresponding to operating conditions. Adsorption rate, particle type, and adsorption capacity are altered between cases.}\label{tab:tab3}
\begin{tabular}{|l| l l l|} 
 \hline
Case & $k$ & Particle & Capacity \\ 
 \hline
 1 & 0.0 & Mono & None \\ 
 2 & 0.0 & Poly & None \\
 3 & 5.3e-6 & Mono & Unlimited \\
 4 & 5.3e-6 & Poly & Unlimited \\
 5 & 5.3e-6 & Mono & Limited \\ 
 6 & 5.3e-6 & Poly & Limited \\ 
 \hline
\end{tabular}
\end{table}

\section{Results}\label{numresults}
\subsection{Concentration profiles}
We show a comparison of the concentration profile in the domain between the monodisperse and polydisperse geometries for all cases listed in Table \ref{tab:tab3}.
\begin{figure}[ht!]
    \centering
    \includegraphics[width=0.8\textwidth]{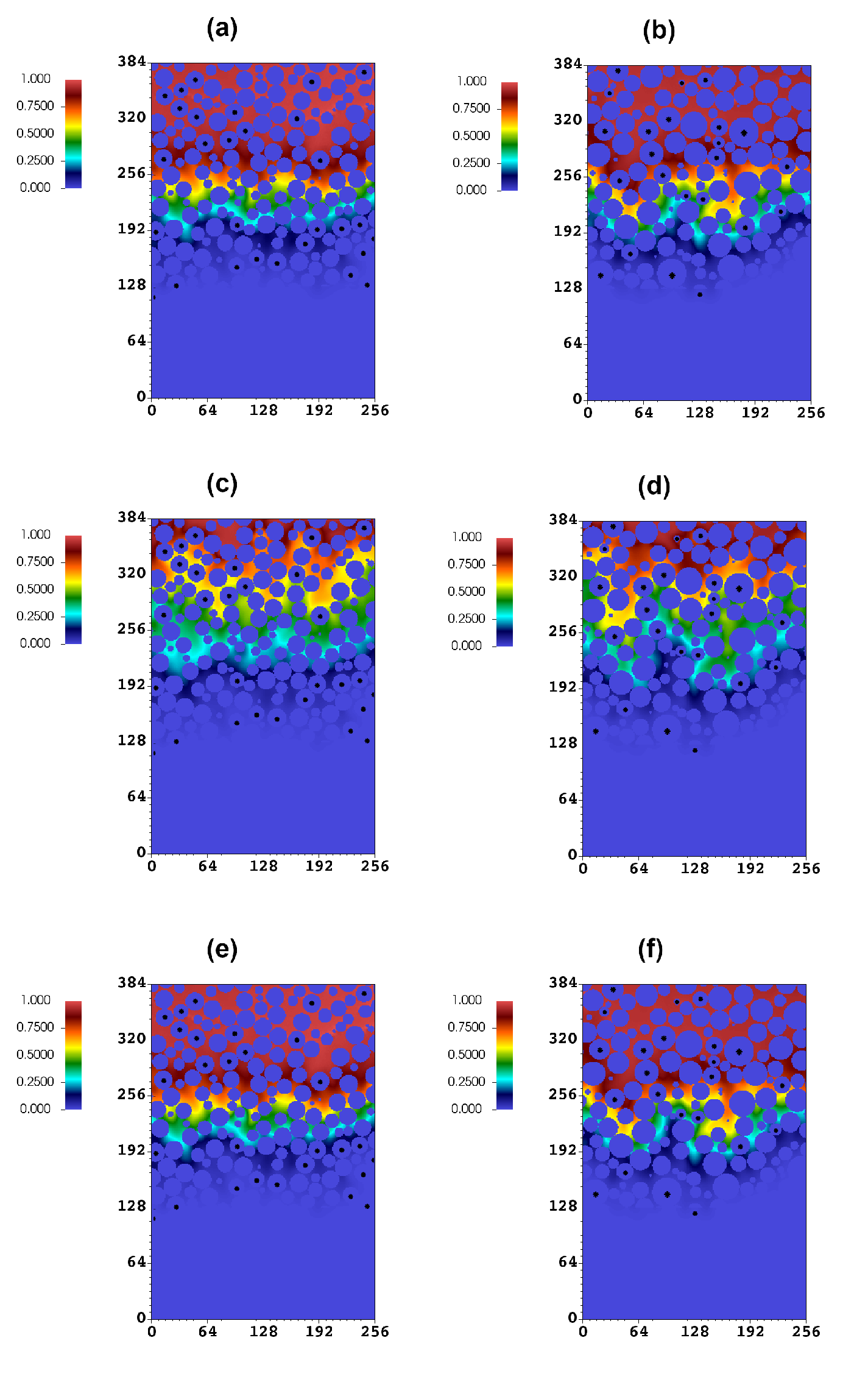}
    \caption{Concentration profiles cut at $x/l_0=140/256$ and $t^* = 0.6 \approx 20$ min. Black stars indicate reactive particles. No capacity: (a) Case 1, (b) Case 2. Unlimited capacity: (c) Case 3, (d) Case 4. Limited capacity: (e) Case 5, (f) Case 6. The polydisperse cases (right panels) appear to generate a less homogeneous concentration front.}\label{fig:fig6}
\end{figure}

Transverse cuts at $x/l_0=140/256$ are shown at the same timestep $t^*=0.6$, given by $t^* = tU/l_0$, which corresponds to the time at which the majority of the reactive particles begin to reach their full capacity, in the cases with limited capacity. Particles marked with black stars indicate reactive particles. Figures \ref{fig:fig6}(a), (c), and (e) display a more homogeneous front, all with the monodisperse geometry. We do see decreased homogeneity in figures \ref{fig:fig6}(b), (d), and (f), where we have the polydisperse geometry. One can see clearly the difference in front homogeneity and indeed the formation of tendrils of higher concentration within these figures. We also observe a general trend of decreased homogeneity in the polydisperse cases with adsorption, (d) and (f), which we investigate in more detail in subsequent sections.

\subsection{Average surface concentration and flux ratio comparison}
\begin{figure}[ht!]
\centering
    \includegraphics[width=\textwidth]{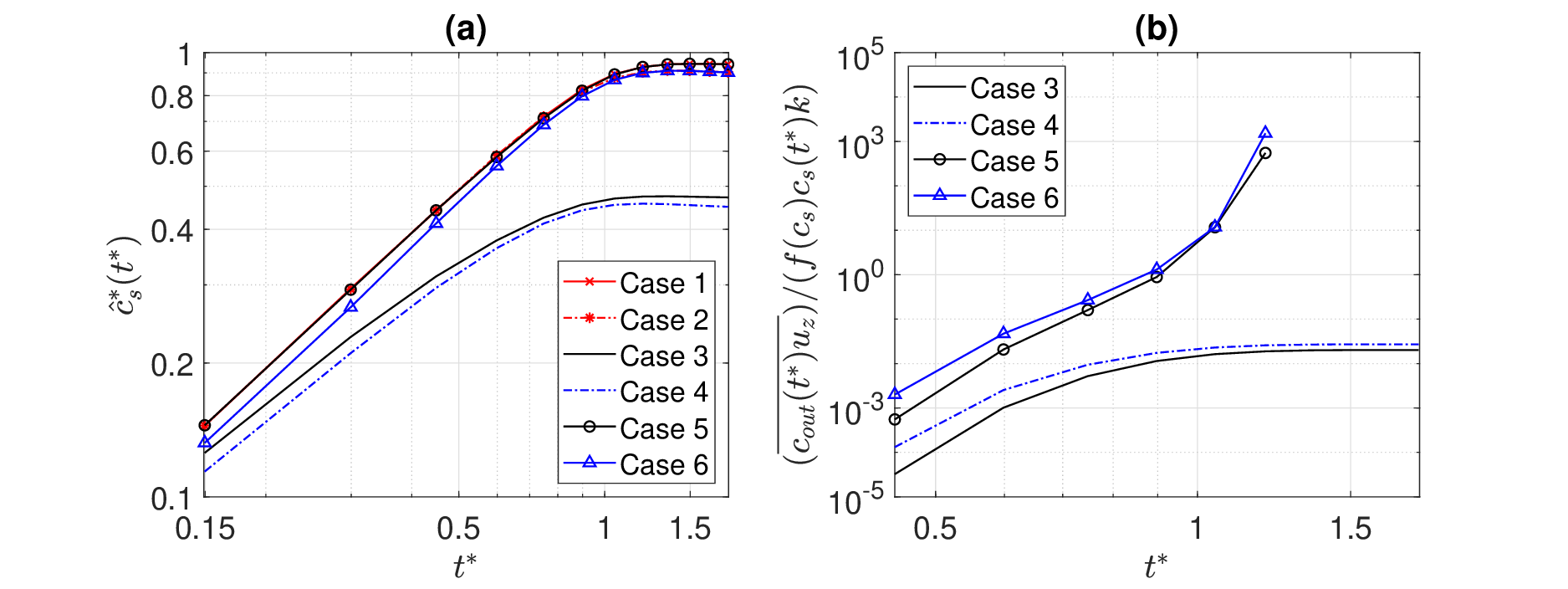}
    \caption{(a) Average surface concentration and (b) breakthrough flux/adsorption flux ratio as a function of dimensionless time. No capacity: Case 1,2. Unlimited capacity: Case 3,4. Limited capacity: Case 5,6}\label{fig:fig7}
\end{figure}

The average surface concentration of all reactive particles is given in Figure \ref{fig:fig7}. This quantity is calculated by summing the concentration on all particle surface nodes and dividing by the total amount of such nodes.
In Figure \ref{fig:fig7}(a) there is a great similarity of behavior between the cases with no capacity and those with a limited adsorption capacity. We see the surface concentrations increase to near unity as the reactive particles are no longer capable of removing more solute from the liquid. On the other hand, if one examines the cases with unlimited capacity we see a significantly reduced surface concentration, around 0.5. This is their effective equilibrium state and further reduction or buildup will not occur without external stimulus. We also note that there is little difference between the monodisperse and polydisperse geometries, implying a similar total reactive surface area despite the difference in individual particle surface areas, which is the case. 

In Figure \ref{fig:fig7}(b) the ratio of the breakthrough flux and the adsorption flux is shown. This quantity can be seen as a measure of the medium leakage reduction as a consequence of biochar adsorption. Note that the cases with the limited capacity model end prematurely due to the particles ceasing to adsorb after reaching their adsorption capacity. The ratio in the beginning of the simulations is extremely small, indicating adsorption is the dominant flux in the system, however this reduces significantly as the simulations progress. This is expected since the concentration flux at the outlet will increase over time as the solute travels through the medium. At longer times the ratio for the unlimited capacity cases stabilizes at around $10^{-2}$ indicating an equilibrium between the adsorption and the breakthrough flux throughout the simulation lifetime. This is not true in the cases with limited adsorption capacity, wherein we see an inflection point at which the particles cease to react and the breakthrough flux quickly becomes the only contributing factor. This period of approaching particle inertness is quite similar for both geometries, indicating a negligible effect of the front homogeneity on this macroscopic quantity.

\subsection{Breakthrough curves (vs. experiments)}
\begin{figure}[ht!]
\centering
    \includegraphics[width=\textwidth]{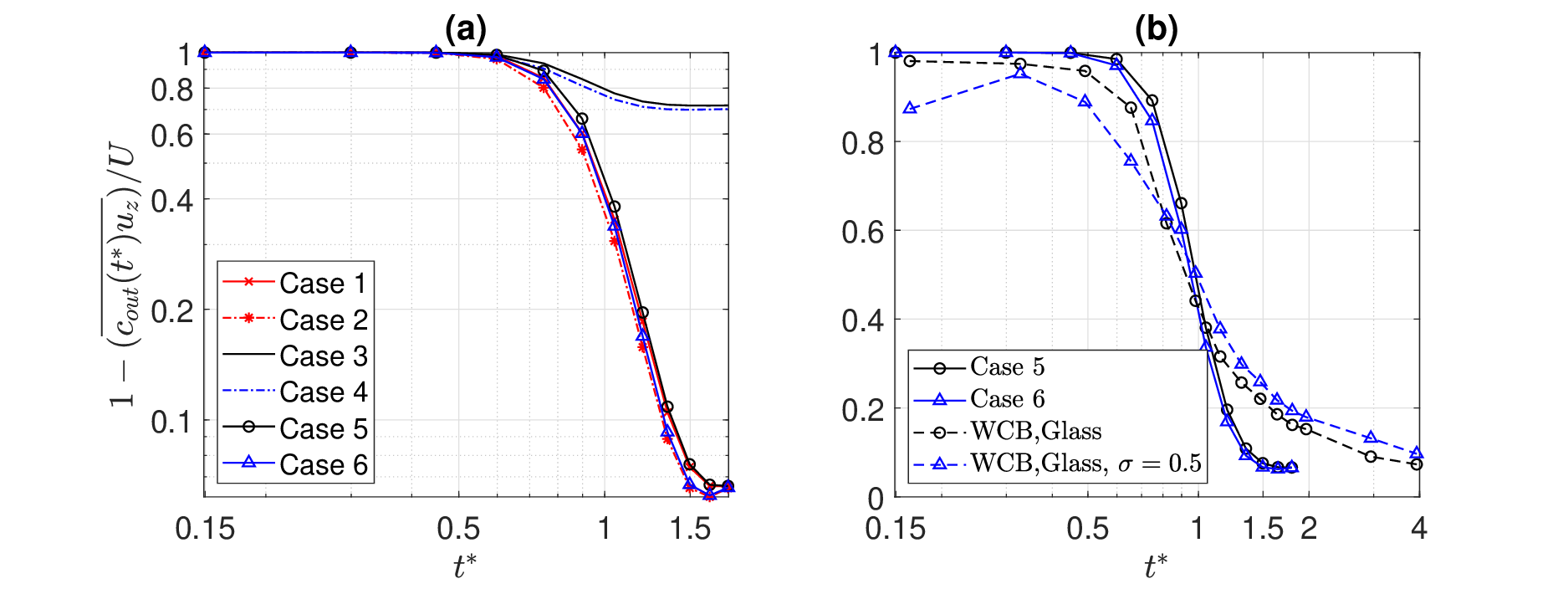}
    \caption{Breakthrough curves for (a) simulation and (b) between like experimental cases and simulations. 1 - no solute reaches outlet, 0 - no adsorption has occurred, all solute reaches outlet. No capacity: Case 1,2. Unlimited capacity: Case 3,4. Limited capacity: Case 5,6}\label{fig:fig8}
\end{figure}

Breakthrough curves are calculated for all cases and displayed in Figure \ref{fig:fig8} where (a) is the average concentration present in a small outlet volume the thickness of one particle diameter calculated as 
\begin{equation}
    \overline{c_{out}(t^*)u_z} = \left(\int_{V_{out}} c_{out}(t^*)u_z \text{d}V_{out} \right) /V_{out}
\end{equation}
where $V_{out} = l_0^2\varepsilon d$ and (b) illustrates the comparison of like cases between the experimental results and the simulations.
In these graphs, a value of 1 represents a full reaction wherein nothing reaches the outlet and 0 represents no change in concentration between inlet and outlet, i.e. nothing has been removed. It is immediately clear in Figure \ref{fig:fig8}(a) that unlimited capacity cases perform significantly better in terms of leakage than those which have a limit or have no reactivity. Interestingly, there is little difference in performance between the cases with no adsorption capacity and those with limited capacity. This is due to the fact that the limited capacity particles reach capacity very swiftly and thus contribute very little to removing the solute after the earlier stages of the simulation. 

This hypothesis is given some weight by Figure \ref{fig:fig8}(b) where two experimental cases are directly compared to the simulation cases with limited adsorption capacity. We find a very similar behavior in that both experiments and simulations swiftly reach capacity and subsequently remove little of the solute. Note that there is an initial period of time for the MB solution to reach the porous media through the piping and we do not take wall effects into account in the simulation; both of which are present in the experiment. Consequently, the experimental curves have a lower gradient and take additional time to reach full capacity, however the steady-state behavior is correctly reflected by the simulations. 

At lower rainfall intensities we expect to see an increased residence time of the adsorbent near the biochar surfaces, however this will not matter if the biochar is already at capacity. In addition to the uncertain reaction kinetics, lower rainfall intensities may also result in unsaturated flow conditions, radically altering the system and introducing capillary forces, making any prediction based upon the current work void. Finally, a lower flow velocity may transition from more complex mixing caused by the microstructure to a more simplified laminar flow in larger pore spaces. 

\subsection{Flow homogeneity and outlet flow profile}
We examine the probability distribution for the streamwise and transverse velocity magnitude as a quantification of the flow homogeneity within each geometry as well as compare between the two. 
\begin{figure}[ht!]
    \centering
    \includegraphics[width=\textwidth]{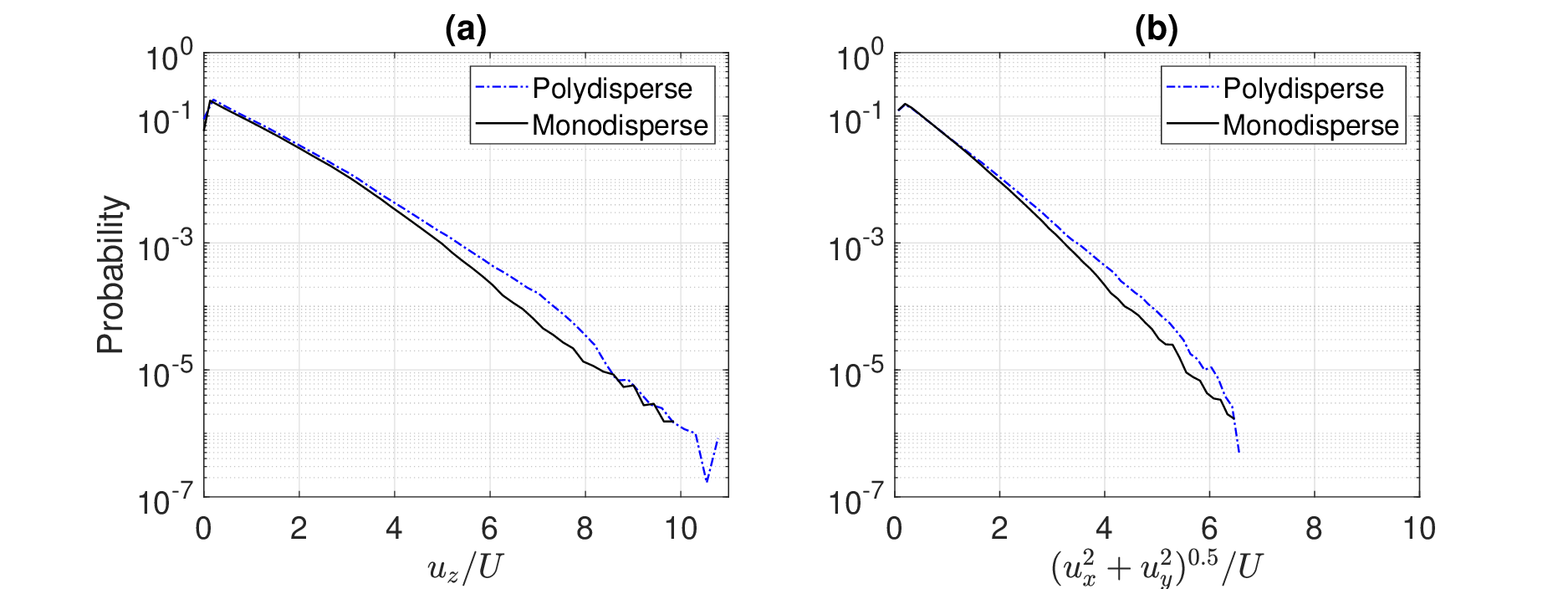}
    \caption{Probability distribution of the domain velocity with (a) $u_z/U$ where $u_z$ is the streamwise velocity component and (b) $(u_x^2 + u_y^2)^{0.5}/U$ the transverse velocity magnitude}\label{fig:fig9}
\end{figure}

Figure \ref{fig:fig9}(a) shows the normalized streamwise velocity and it is clear that the streamwise velocity is largely similar to that of the mean velocity. However, variations up to 10 times that value exist within the flow field which is also reflected in the distribution of the transverse velocity. There is no real discernible difference in the distributions between the two geometries, thus we can be reasonably confident that the flow alone is not a primary factor in notable differences observed between cases. Indeed, we do see increased inhomogeneity induced by increased adsorption in the front profiles shown in Figure \ref{fig:fig6}. 

\begin{figure}[ht!]
\centering
    \includegraphics[width=\textwidth]{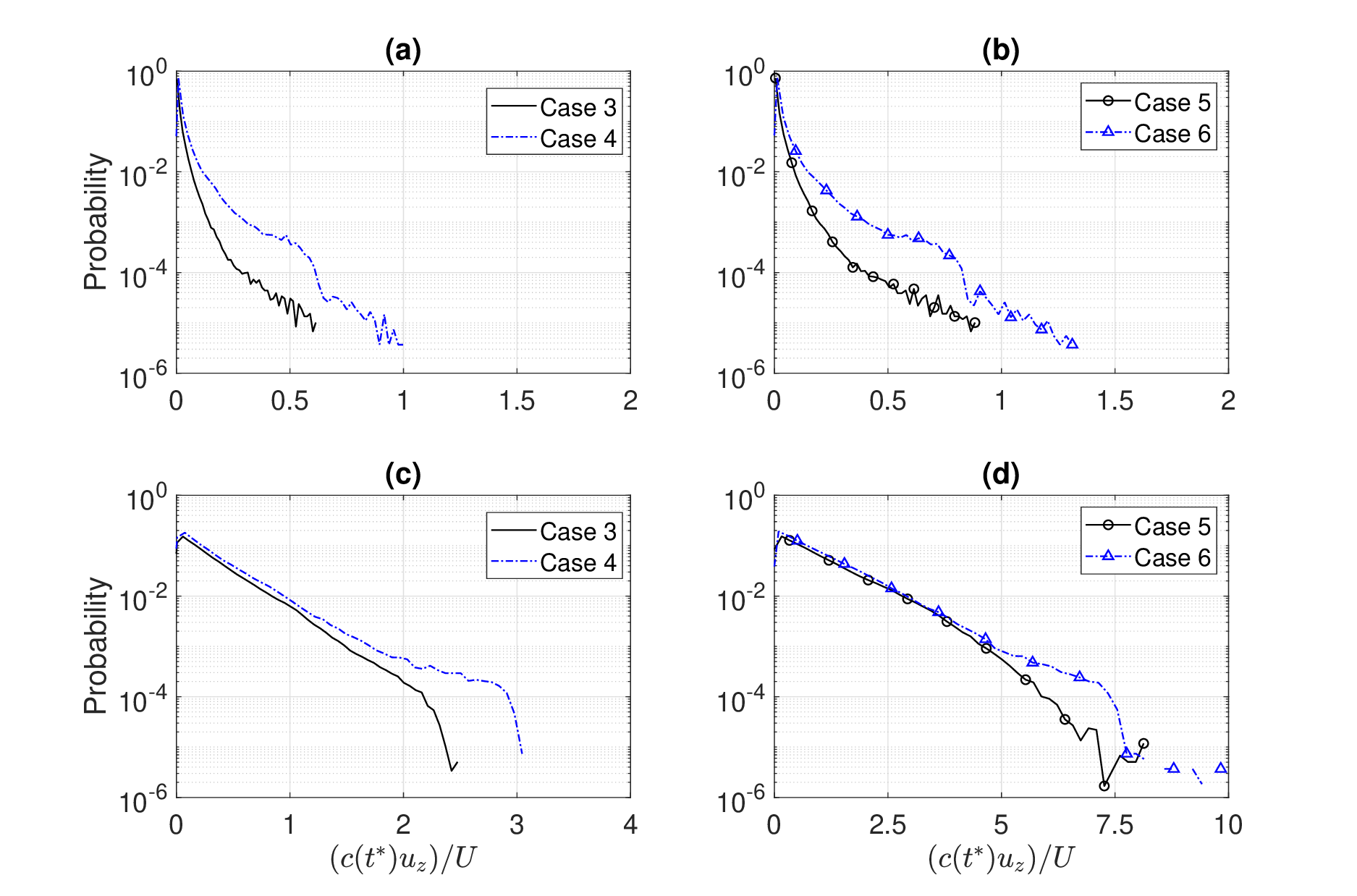}
    \caption{Breakthrough flux distributions for (a) Unlimited capacity, $t^*=0.6$ (b) limited capacity, $t^*=0.6$ (c) Unlimited capacity, $t^*=1.5$ (d) limited capacity, $t^*=1.5$}\label{fig:fig10}
\end{figure}

Figure \ref{fig:fig10} displays the probability distributions for the breakthrough flux, measured $c(t^*)u_z/U$, as a quantification of the flow homogeneity at the outlet. This gives an indication of how the flow interacts with the reactive particles, producing tendrils of increased concentration as opposed to a more uniformly distributed concentration front. 

We compare the distributions during the period where the solute first reaches the outlet ($t^*=0.6$), figures \ref{fig:fig10}(a) (unlimited capacity) and \ref{fig:fig10}(b) (limited capacity) and when the steady state at the outlet is reached ($t^*=1.5$), figures \ref{fig:fig10}(c) (unlimited) and \ref{fig:fig10}(d) (limited). A value near 1 indicates tendrils of high concentration transported at a velocity around that of the streamwise mean velocity. Deviations from this imply a change to the local concentration or streamwise velocity. In other words, fluctuations around the mean of the concentration flux. Note a strong similarity in distribution between \ref{fig:fig10}(a) and \ref{fig:fig10}(b), though the measured concentration flux is slightly higher in the limited capacity case. In all cases the monodisperse geometry exhibits a lower and tighter distribution of outflow concentration, indicating higher and more efficient adsorption of the solute. The largest deviation is found in Figure \ref{fig:fig10}(d), which compared to the unlimited capacity case in \ref{fig:fig10}(c), has a higher concentration flux magnitude. We know from our analysis of the flow itself there is little influence on the flow induced by each geometry, thus the adsorption rate spatial distribution is the likely cause for this phenomenon. 

\subsection{Particle adsorption capacity}
\begin{figure}[ht!]
\centering
    \includegraphics[width=\textwidth]{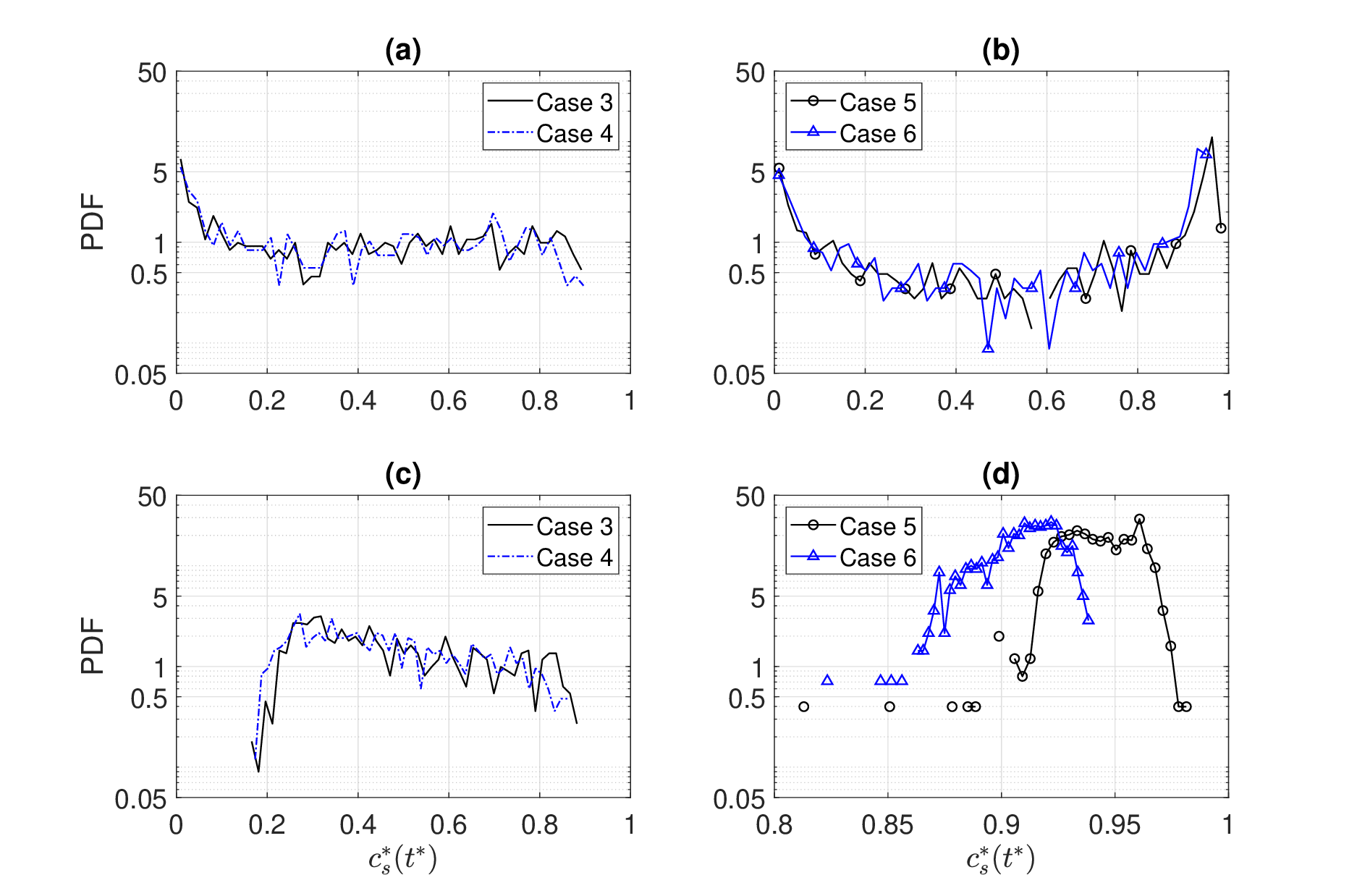}
    \caption{Particle average surface concentration PDFs for (a),(b) $t^*=0.6$ and (c),(d) $t^*=1.5$. Unlimited capacity: Case 3,4. Limited capacity: Case 5,6}\label{fig:fig11}
\end{figure}

\begin{figure}[ht!]
\centering
    \includegraphics[width=\textwidth]{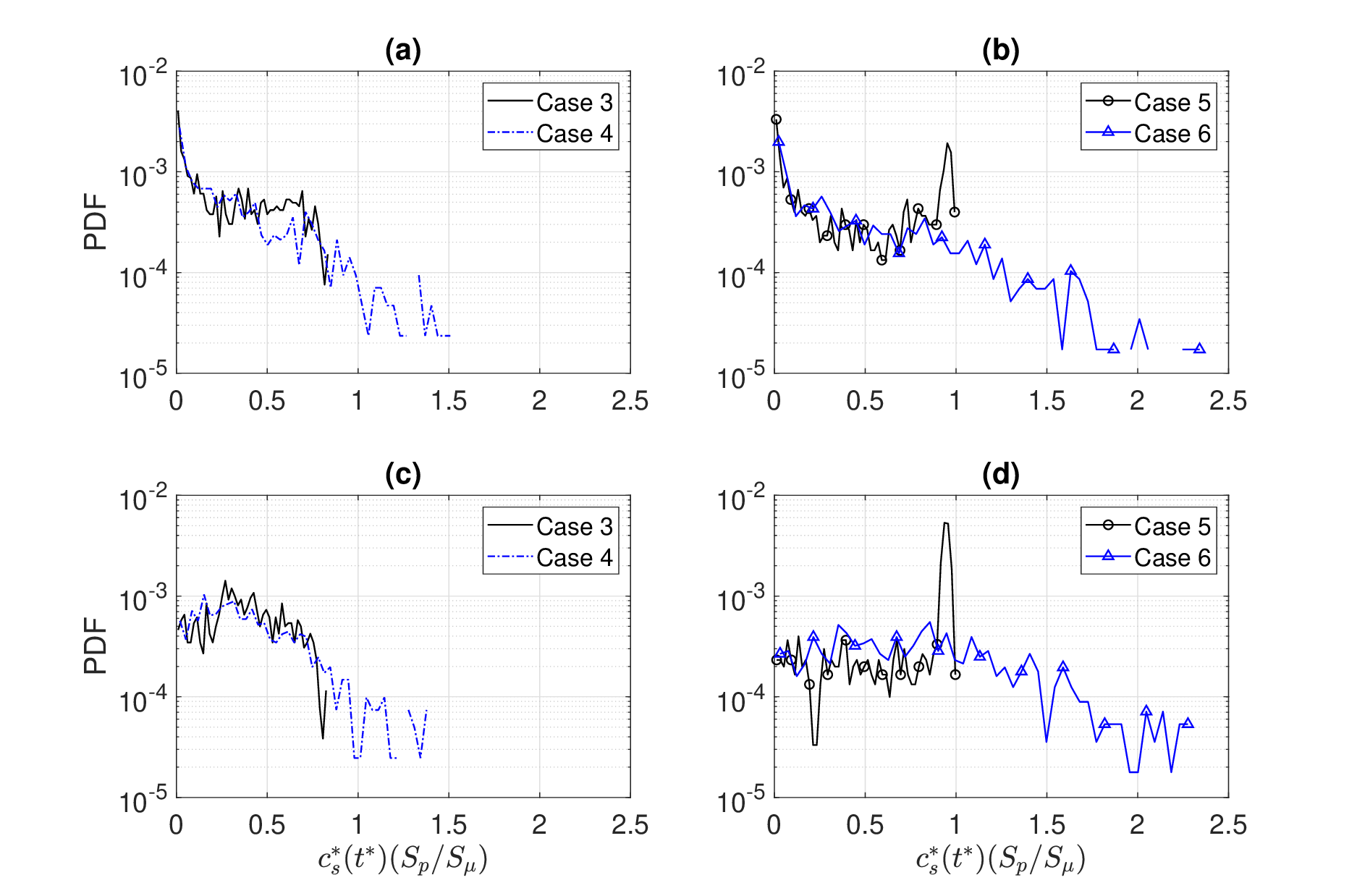}
    \caption{Surface concentration flux PDFs for (a),(b) $t^*=0.6$ and (c),(d) $t^*=1.5$. Unlimited capacity: Case 3,4. Limited capacity: Case 5,6}\label{fig:fig12}
\end{figure}

Figure \ref{fig:fig11} displays the PDFs of average particle surface concentration $c^*_s$ during the transition stage $t^*=0.6$, figures \ref{fig:fig11}(a),(b) wherein particles are reaching their capacity and begin to become non-reactive. The lower figures, (c),(d) show the same cases, however the timestep is altered to reflect the steady-state distribution found at $t^* = 1.5$. Figures \ref{fig:fig11}(a),(c) are unlimited capacity and (b),(d) are limited capacity.  The order of cases and times is replicated for Figure \ref{fig:fig12}.

Figure \ref{fig:fig12} displays PDFs for the surface concentration flux $c^*_sS_p/S_\mu$, using the prefactor $S_p/S_\mu$, where $S_p$ is individual particle surface area and $S_\mu$ is mean particle surface area. This ratio is the measure of the available adsorptive surface area compared to that of the ideal surface area of a representative particle. When combined with a particle's surface concentration we can get an idea of which particles are more effective individually in terms of the adsorptive flux.   

In Figure \ref{fig:fig11} we show the distribution of each particles' average surface concentration, taken at an intermediate and steady-state time. We observe no notable differences between the monodisperse and polydisperse geometries. At the intermediate time the distribution for the limited capacity cases exhibits the U-shaped distribution typical of diffusion-dominated systems, with two peaks at the extrema. In the cases with unlimited reactivity the peak at higher concentration disappears due to the continuous adsorption, figure \ref{fig:fig11}(a). At the steady-state, the cases with limited adsorption tend towards concentration of 1 with a very narrow distribution of values whereas the unlimited are more widely distributed, figures \ref{fig:fig11}(c),(d). 

When one examines the surface concentration flux $c^*_sS_p/S_\mu$ in the cases with unlimited capacity, figures \ref{fig:fig12}(a),(c), we observe a similar behavior to that of the average particle surface concentration. As expected, a similar trend is also evident in the monodisperse case with limited capacity, figures \ref{fig:fig12}(b),(d), when compared to the distribution evolution in time of figures \ref{fig:fig11}(b),(d). Indeed $S_p/S_\mu \approx 1$ in the monodisperse case.

This is not reflected in the polydisperse case with limited capacity. At longer times where the concentrations are uniformly $c^*_s \approx 1$, the fluxes will tend to the distribution of $S_p/S_\mu$, figure \ref{fig:fig12}(d). Due to the fact that the particle radii are normally distributed, the probability of $S_p/S_\mu$ will be $\chi^2$ distributed, exhibiting an exponential decay away from the mean. 

\begin{table}[ht!]
	\centering
		\caption{Ratio of mean particle size with concentration higher than concentration threshold vs. mean particle size of the rest at $t^*=0.6$}\label{tab:tab4}
\begin{tabular}{|l| c c c c|} 
 \hline
 Concentration & 0.3000 & 0.5000 & 0.7000 & 0.9000\\ 
 \hline
 Ratio of means & 0.9924 & 0.9908 & 0.9515 & 0.8853\\ 
 \hline
\end{tabular}
\end{table}
Interestingly, at the intermediate time, the distribution of the fluxes preserves the peak at low concentration whereas the peak at higher concentration visible in the monodisperse case here is absent for the polydisperse case. We calculate the ratio of mean particle size with concentration higher than a certain threshold against the mean size of the remainder particles, shown in Table \ref{tab:tab4}. It is noticeable that high surface concentrations, that is above a specified threshold, are found on smaller particles. In turn the distribution of fluxes ($c^*_sS_p/S_\mu$) decreases uniformly as the flux itself increases, figure \ref{fig:fig12}(b). This may indicate that some particles in the polydisperse geometry are being underutilized. If one examines the cuts shown in Figure \ref{fig:fig6} where the polydisperse cases display preferential pathing we see some occurrences of preferential paths probably following high concentrations around smaller particles which have reached capacity. 

The behavior discussed above regarding the monodisperse geometry is indicative of a more homogeneous front, where the particles are either being approached by the front or are surrounded by it, whereas the front reaches the particles in the polydisperse geometry at different times, leading to a higher distribution of different magnitudes. We note that the monodisperse case exhibits an overall higher surface concentration at the steady-state than that of the polydisperse cases. This is coupled with a similarly lower breakthrough flux. One explanation for this is the increased inhomogeneity observed in the concentration profiles for the polydisperse geometry. This preferential pathing by the formation of tendrils of higher concentration removes solute from reaching particles that would otherwise aid in the removal of concentration from the flow. This channeling effect reduces the efficiency of the reactive particles as some available surface area is left unused, effectively underutilizing them. A more homogenenous front allows for the maximum surface area to be reached at any given time.

\begin{figure}[ht!]
    \centering
    \includegraphics[width=\textwidth]{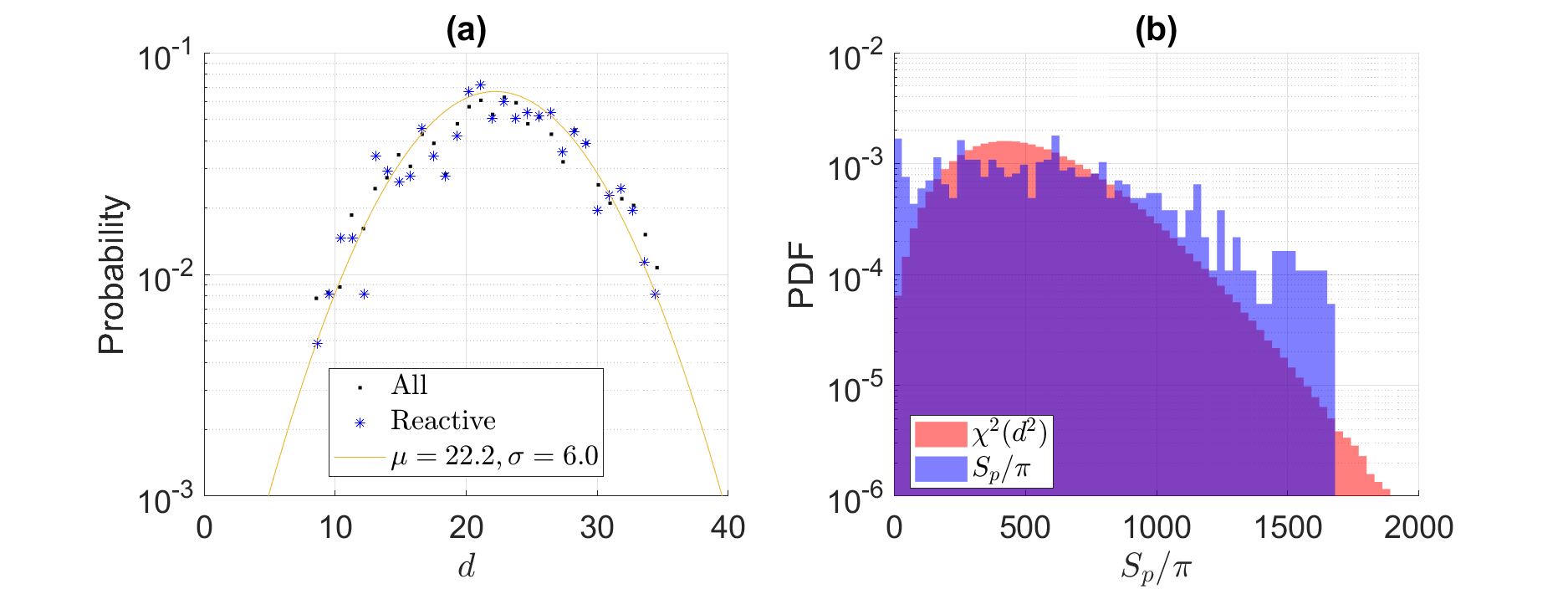}
    \caption{(a) Particle diameter $d$ distribution in l.u.. Dots indicate all domain particles, stars are just reactive particles. Line is the normal distribution function with $\mu=22.2,\sigma=6.0$. (b) PDF of reactive particle surface area $S_p/\pi$ (blue), $\chi^2(d^2)$ distribution of particle diameter $d$ (red)}\label{fig:fig13}
\end{figure}

Figure \ref{fig:fig13} shows the particle diameter and surface area distributions to aid in the interpretation of the results as discussed above.
Figure \ref{fig:fig13}(a) shows the distribution of reactive particles only and all particles for the polydisperse case as well as the resultant normal distribution curve. The distribution of dimensionless surface area is shown in Figure \ref{fig:fig13}(b) for the polydisperse case only. It is compared to the ideal PDF of diameter squared $d^2$ which follows a $\chi^2$ distribution.  The difference observed are due to surface contact or cutting along the domain edges. Despite this fact, we are satisfied that the distributions shown here justify the discussion we have entertained regarding geometry effects and the potential for preferential pathing induced by the polydisperse geometry in particular.

We can now state with reasonable certainty that reactive particles do influence the concentration profiles, particularly the available reactive surface area. An advantage may be achieved by using more mondisperse particles, as they promote a more homogeneous front for the advancing solute and reduce the solute leakage, particularly in the case of particles with limited adsorbing capacity.

\subsection{Modifying the reaction kinetics}
We close the paper with a short discussion on the practicalities of accurately modelling the adsorption dynamics present in reality on the surface of biochar. There are several different mechanisms by which adsorption occurs; namely pore-filling, hydrophobic interaction, surface complexation, electrostatic interaction, hydrogen bonding interaction, $\pi-\pi$ electron-donor-acceptor (EDA) interaction, and intra-particle diffusion. In addition it has been shown that thermal conditions as well as the pH of the environment play a non-negligible role in the reaction speed \cite{Ji2024,Gupta2020}, as does aging or the presence of dissolved organic matter. The sum total of considering individual biochar characteristics, their intrinsic modes of adsorption, as well as environmental factors renders the modelling a significant challenge. In general current works have made use of first and second order pseudo-kinetic models alongside Langmuir and Freundlich isotherms. Some forays have been undertaken in the realm of applied artificial neural networks and statistical chemical modelling however this is still in its infancy for such an application. If one wishes to enhance the models governing the reaction kinetics we need more experimental data to determine which of the above modes of adsorption should be considered and how the environmental and aging aspects can be included in the models. While some kind of pre-factors can be easily applied they are not representative of any physical phenomenon and some possible empirical model msut be constructed which accurately captures these effects may be a simple and efficient solution. Alternatively the full chemical species and reactions can be included in the modeling but this undertaking may not be compatible with how the lattice Boltzmann method operates but would ensure a more chemically consistent interaction. Pore filling and intra-particle diffusion might be handled via some volumetric functions that account for internal pore volume and pore network but this is also impractical to implement. Whichever strategy is used, more experimental data is likely to be required and understanding the fundamental physical interactions in such a system are a good place to aid in determining where to next add the increased realism required for more accurate and applicable predictions. 

\section{Conclusion}\label{conclusion}
Experiments have been undertaken to determine the adsorption rate and capacity of two different kinds of biochar alongside breakthrough curves for same. The breakthrough curves are calculated using the absorbance of an MB solution over a period of time. Simulations are run using the experimental data and a comparison is made between the predicted breakthrough curves and the experimental results, with good agreement found. 

A deeper investigation of the underlying factors determining the concentration field evolution is described. This analysis consists of flow velocity and breakthrough concentration profiles, average and individual surface concentration distributions, particle capacity distributions,  and a flux ratio comparison. The results show that the flow velocity distribution remains the same for both simulated geometries, implying that the differences in concentration distribution may not be a result of the flow velocity profile. Examination of the breakthrough curves and individual particle concentration distributions show that the polydisperse geometry promotes preferential pathing throughout the medium and leads to a more varied distribution of particle states than those found in the monodisperse cases. This preferential pathing is undesirable in the context of removal of contaminants from the flow, as a more homogeneous front ensures the randomly distributed reactive particles will interact with the solute, whereas tendrils may bypass them entirely and flow directly for the outlet.

We have chosen to use spherical packed beds over a realistic soil substrate geometry due to the intrinsic inhomogeneity found within soil scans taken using x-ray microtomography (XMT) or any similar technique. A chosen domain may contain few very large particles and multitudes of tiny particles or anywhere in between, rendering the porosity of any such domains fluctuating wildly, making comparisons difficult. In future it is worthwhile to determine if the trends observed in this work hold true for the realistic soils as well.

If one extends these results to application recommendations, we can state that the observed differences between the mono- and polydisperse cases are larger than those observed when biochar is present or not. This would correspond to cases 1 and 2 (no biochar) and 5 and 6 (biochar with a limited capacity). The concept of limited adsorption capacity is analogous to the realistic adsorbance occurring in the experimental biochar, which cannot remove the solute indefinitely. These cases only slightly outperform the cases in which there are no reactive particles present. This is due to the rate at which the limited capacity particles reach capacity and become inert. Cases 3 and 4 have unlimited capacity, which one could say is analogous with roots or some other living biological matter that does not have a capacity in the same manner as biochar. 

The practical implication here is that monodisperse beds are preferred as they promote more even distribution of the solute within the domain and thus allow the random adsorbers to act with maximum effect, i.e. reaction-limited. Additionally, the presence of adsorbers with an unlimited adsorption capacity must be taken into account as this can dramatically effect the results, particularly in proximity to each other. This effect will obviously increase if the adsorption rate is increased.

If one wishes to increase the efficacy of biochar, a high absorptive rate with a similarly high capacity will greatly reduce breakthrough. This can be accomplished by modifying the material itself and/or increasing the total quantity of biochar within the soil, which will consequently increase its capacity. Current practice suggests no more than 15\% biochar by volume is beneficial given common rainfall quantities; however our results show that performance can be increased if one takes into account the factors mentioned above, with the additional mention of total reactive surface area and the idea of induced preferential pathing. Lower rainfall intensities will cause higher residence time of the concentration front near the biochar; however once they saturate, which we show can take place quite quickly, breakthrough will still occur. Thus a more accurate method for modeling the capacity of different adsorbers should be investigated to improve the predictive ability of the models. Finally, given that monodisperse particles promote a more homogeneous concentration front and that total reactive area matters more than individual particle size we suggest that more uniform sized substrates are used.

\backmatter

\bmhead{Acknowledgements}

This work was supported by the Swedish Research Council for Environment, Agricultural Sciences and Spatial Planning (FORMAS), Grant No. 2019–01261. The computations were enabled by resources provided by the Swedish National Infrastructure for Computing (SNIC), partially funded by the Swedish Research Council through Grant Agreement No. 2018–05973.

\bibliography{biblio_Pettersson2024v2}% common bib file

\section*{Declarations}
\begin{itemize}
\item Funding: FORMAS Grant No. 2019–01261\\
\item Competing Interests: The authors have no competing interests to declare that are relevant to the content of this article.\\
\item Ethics approval and consent to participate: Not applicable\\
\item Consent for publication: Not applicable\\
\item Data availability: The datasets generated during and/or analysed during the current study are available from the corresponding author on reasonable request.\\
\item Materials availability: Not applicable\\
\item Code availability: Not applicable \\
\item Author contribution: Conceptualization: Dario Maggiolo, Oskar Modin, Angela Sasic Kalagasidis; Formal Analysis: Kaj Pettersson, Albin Nordlander, Dario Maggiolo; Funding acquisition: Dario Maggiolo; Investigation: Kaj Pettersson, Albin Nordlander; Methodology: Dario Maggiolo, Oskar Modin; Project administration: Dario Maggiolo; Resources: Dario Maggiolo, Angela Sasic Kalagasidis, Oskar Modin; Software: Kaj Pettersson, Dario Maggiolo, Albin Nordlander; Supervision: Dario Maggiolo, Oskar Modin; Validation: Kaj Pettersson, Albin Nordlander; Visualization: Kaj Pettersson, Albin Nordlander; Writing – original draft: Kaj Pettersson; Writing – review \& editing: Kaj Pettersson, Dario Maggiolo, Angela Sasic Kalagasidis, Oskar Modin
\end{itemize}
\noindent
\end{document}